\documentclass{llncs}
\usepackage{url}

\usepackage{graphicx}

\begin{document}
\mainmatter              % start of a contribution
\title{A Generic Modelling Framework\\for Last-Mile Delivery Systems}
\titlerunning{Designing and Implementing a Generic ABM Framework}  % abbreviated title (for running head)
\author{
        \"{O}nder G\"{u}rcan\inst{1}*\orcidID{0000-0001-6982-5658} \and
        Timo Szczepanska\inst{1}\orcidID{0000-0003-2442-8223} \and
        Vanja Falck\inst{1}\orcidID{0000-0003-0855-5943} \and
        Patrycja Antosz\inst{1}\orcidID{0000-0001-6330-1597} \and \\
        Merve Seher Cebeci\inst{2}\orcidID{0000-0002-9248-2128} \and
        Michiel de Bok\inst{2}\orcidID{0000-0002-5391-0652} \and
        Rodrigo Tapia\inst{2}\orcidID{0000-0001-7132-5695} \and
        Lóránt Tavasszy\inst{2}\orcidID{0000-0002-5164-2164}
}
\authorrunning{
    \"{O}. G\"{u}rcan et al.
}
\tocauthor{
        \"{O}nder G\"{u}rcan,
        Timo Szczepanska,
        Vanja Falck,
        Patrycja Antosz,
        Merve Seher Cebeci,
        Michiel de Bok, 
        Rodrigo Tapia and
        Lóránt Tavasszy
}
\institute{
Center for Modeling Social Systems,\\
NORCE Norwegian Research Center AS,\\
Universitetsveien 19, Kristiansand, Norway\\
    %\email{xyzt@norceresearch.no}
\and 
Department of Transport and Planning,\\ 
Faculty of Civil Engineering and Geosciences,\\ 
Delft University of Technology,\\ 
Stevinweg 1, 2628 CN Delft, The Netherlands
}

\maketitle              % typeset the title of the contribution

\begin{abstract}
Large-scale social digital twinning projects are complex with multiple objectives. 
For example, a social digital twinning platform for innovative last-mile delivery solutions may aim to assess consumer delivery method choices within their social environment. 
However, no single tool can achieve all objectives. Different simulators exist for consumer behavior and freight transport. 
Therefore, we propose a high-level architecture and present a blueprint for a generic modelling framework. 
This includes defining modules, input/output data, and interconnections, while addressing data suitability and compatibility risks. 
We demonstrate the framework's effectiveness with two real-world case studies.
\end{abstract}
\section{Introduction}
\label{sec:Introduction}

The rapid growth of urban populations has significantly challenged last-mile transportation systems, particularly in achieving effective, resilient, safe, and sustainable delivery methods \cite{kiba2021sustainable}. In response, digital twinning projects for cities are emerging as comprehensive solutions that integrate various stakeholders, including governments, industry leaders, solution providers, and researchers \cite{Weil2023}. These projects focus on co-developing innovative last-mile delivery solutions through shared space utilization models, creating a multifaceted approach to urban logistics.
Social digital twinning, which creates digital replicas of physical assets and social systems using as latest data as possible, is a powerful tool for managing the complexity of urban logistics. 
By simulating city infrastructure, transportation networks, delivery operations, and human behaviour, social digital twins enable stakeholders to optimize efficiency, reduce emissions, and improve safety. However, due to the complexity of urban logistics, no single simulation tool can address all aspects.

This paper proposes a generic modelling framework that can be calibrated for different cities, functioning as living labs. The framework integrates various simulation models into a unified system adaptable to each city's unique needs. It aims to facilitate collaboration among multidisciplinary partners from diverse countries, enabling the co-creation of innovative, sustainable, and scalable last-mile solutions across urban environments.
We discuss the components and design of this modelling framework, using the collaborative large-scale project URBANE\footnote{Upscaling Innovative Green Urban Logistics Solutions Through Multi-Actor Collaboration and PI-Inspired Last Mile Deliveries, \url{https://www.urbane-horizoneurope.eu}, accessed on 31/08/2024.} as a case study. The framework supports stakeholder collaboration and integrates green automated vehicles and shared space utilization models into last-mile delivery systems. Additionally, we explore its implementation in different cities, highlighting its potential to enhance resilient, safe, and sustainable urban transportation systems.
The contributions of this paper are as follows:

\begin{itemize}
    \item \textbf{Introduction of a Generic Modelling Framework:}  A high-level architectural blueprint for integrating various simulation tools to manage the complexities of urban last-mile delivery (LMD) systems.
    \item \textbf{Incorporation of Real-World Case Studies:} Practical applications of the framework through two case studies, demonstrating its adaptability and effectiveness in diverse urban settings.
    \item \textbf{Focus on Sustainability and Efficiency:} Enhancing LMD systems' sustainability and efficiency through better decision-making and optimized resource allocation.
\end{itemize}

This paper is organised as follows. 
Section \ref{sec:Last-Mile-Delivery-Systems} provides background on LMD systems and their modelling.  
Section \ref{sec:HUMAT} and Section \ref{sec:MASS-GT} describe the simulation tools used.
Section \ref{sec:Modelling-Framework} idetails our integrated modelling framework for simulating consumer decisions in LMD systems. 
Section \ref{sec:Crowdshipping-Service} and Section \ref{sec:Parcel-Locker-Service} demonstrate the framework's effectiveness through two case studies: crowdshipping and parcel locker services.
Finally, Section \ref{sec:Discussion-and-Conclusion} provides an overall discussion and concludes the paper. 

\section{Last-Mile Delivery Systems}
\label{sec:Last-Mile-Delivery-Systems}

Last-mile delivery is the final stage of delivering a package to the end recipient, typically a consumer. The demand varies by type: business-to-consumer (B2C), business-to-business (B2B), or consumer-to-consumer (C2C). 
B2C deliveries, require fast and reliable service, while B2B involves larger, more complex shipments, and C2C in peer-to-peer marketplaces presents unique challenges. Flexible, decentralized solutions are essential to meet these varied demands and optimize logistics for customer satisfaction.
Urban freight transport, crucial for last-mile logistics, differs from broader systems due to shorter distances, varied commodities, and urban constraints. Modeling last-mile delivery is complex, involving both descriptive and prescriptive approaches \cite{tavasszy2023overview}. 

Descriptive models aim to understand the dynamics of goods delivery in cities, assessing impacts on the economy, environment, and society, and may involve stakeholder analysis \cite{kiba2021sustainable} and understanding the interrelation between last-mile logistics entities  \cite{harrington2016identifying}.
Prescriptive models focus on designing efficient urban LMD networks using operations research, often combining optimization and simulation techniques.

A key challenge in LMD models is ensuring reproducibility across diverse contexts. ABM offers a promising approach that can overcome the limitations of traditional methods in LMD systems \cite{macal2016everything}. While ABMs can address urban freight transport issues \cite{anand2015agent}, achieving the right balance of precision and generalizability to fully capture stakeholder behavior is difficult. Many models fail to account for interdependencies among activities and processes due to inadequate time dependence and limited representation of social dynamics in communities and business ecosystems \cite{Cebeci2023,zenezini2018new}.

\section{Simulation Tool: HUMAT}
\label{sec:HUMAT}

HUMAT is a socio-cognitive agent-based modeling architecture that creates artificial populations where individuals have dynamic beliefs and social networks, influencing their decisions and communication. 
Integrating cognitive, social interaction, and network theories, HUMAT links motivations with decision-making and networked communication. 
Initially developed for simulating urban innovation adoption \cite{Antosz2019}, HUMAT has been applied to various scenarios, including traffic management, sustainable energy transitions, and stress-related decision-making \cite{Antosz2023}. 
Recently, it was used to study opinion dynamics during COVID-19 \cite{Li2023} and integrated into the URBANE project as a platform-independent variant \cite{Gurcan2023}.

\begin{figure}[h]
\centering
\includegraphics[width=0.80\textwidth]{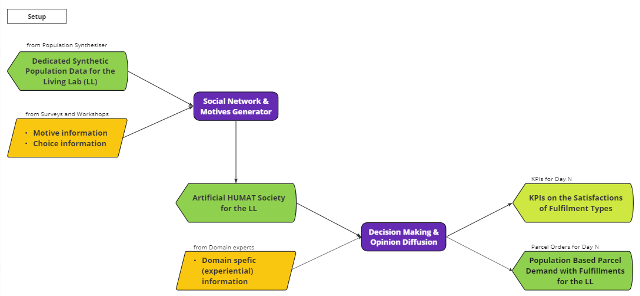}
\caption{HUMAT - Setup phase}
\label{fig:HUMAT-Setup-Phase}
\end{figure}

The HUMAT model is set up in two steps (Figure \ref{fig:HUMAT-Setup-Phase}). First, the Social Network \& Motive Generator module processes synthetic population data and information on motives and choices to create HUMAT agents, each with distinct internal motives. It then constructs three types of homophily networks: friendship, job-related, and neighborhood networks.
In the second step, the Decision Making \& Opinion Diffusion module uses the artificial HUMAT society and case-specific behavioral data to manage the agents' decision-making processes. It outputs initial satisfaction levels for different parcel fulfillment choices and generates KPIs summarizing these levels for various population subgroups.

\section{Simulation Tool: MASS-GT}
\label{sec:MASS-GT}

\begin{figure}[h]
\centering
\includegraphics[width=0.80\textwidth]{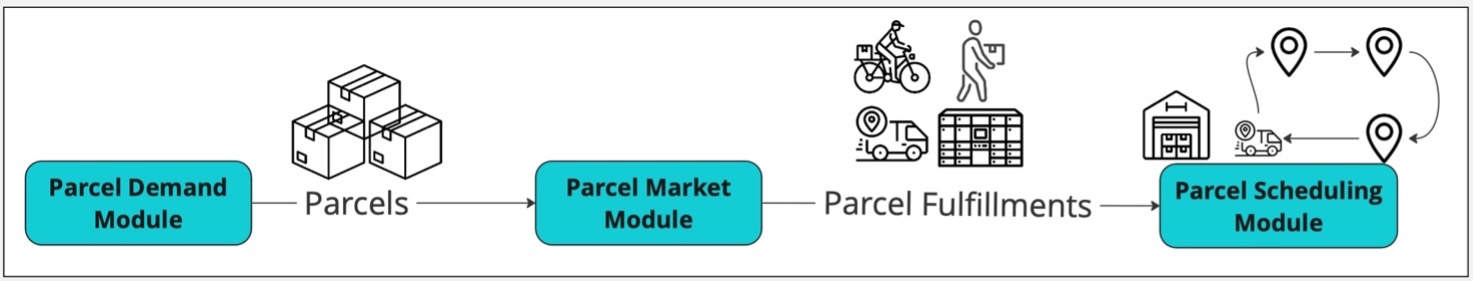}
\caption{MASS-GT modules}
\label{fig:MASS-GT-Modules}
\end{figure}

The Multi-Agent Simulation System for Goods Transport (MASS-GT) is an ABM focused on freight transport, involving various agents like producers, shippers, carriers, and policymakers \cite{deBok2018}. In the URBANE project, MASS-GT is used to simulate the LMD market, with emphasis on three key modules: Parcel Demand, Parcel Market, and Parcel Scheduling (Figure \ref{fig:MASS-GT-Modules}).

\textit{Parcel Demand} generates parcel demand using population indicators, sociodemographic data, and contextual data like network structure. Demand is simulated through Monte Carlo methods, and parcels are allocated to carriers based on market share and delivery success rates, producing total parcel numbers and KPIs.
\textit{Parcel Market} assigns carriers to fulfill parcels, generating delivery trips using methods like parcel lockers, crowdshipping, or traditional couriers, and creates indicators such as extra trips or capacity usage.
\textit{Parcel Scheduling} handles parcels not assigned in the Parcel Market, generating carrier tours and calculating tour distances for traditional couriers.

\section{Proposed Generic Modelling Framework}
\label{sec:Modelling-Framework}

%This section introduces an integrated ABM framework that simulates how consumers make and evolve their delivery decisions within their social environment, considering various LMD options and the role of logistics service providers.

\begin{figure}[h]
\centering
\includegraphics[width=0.82\textwidth]{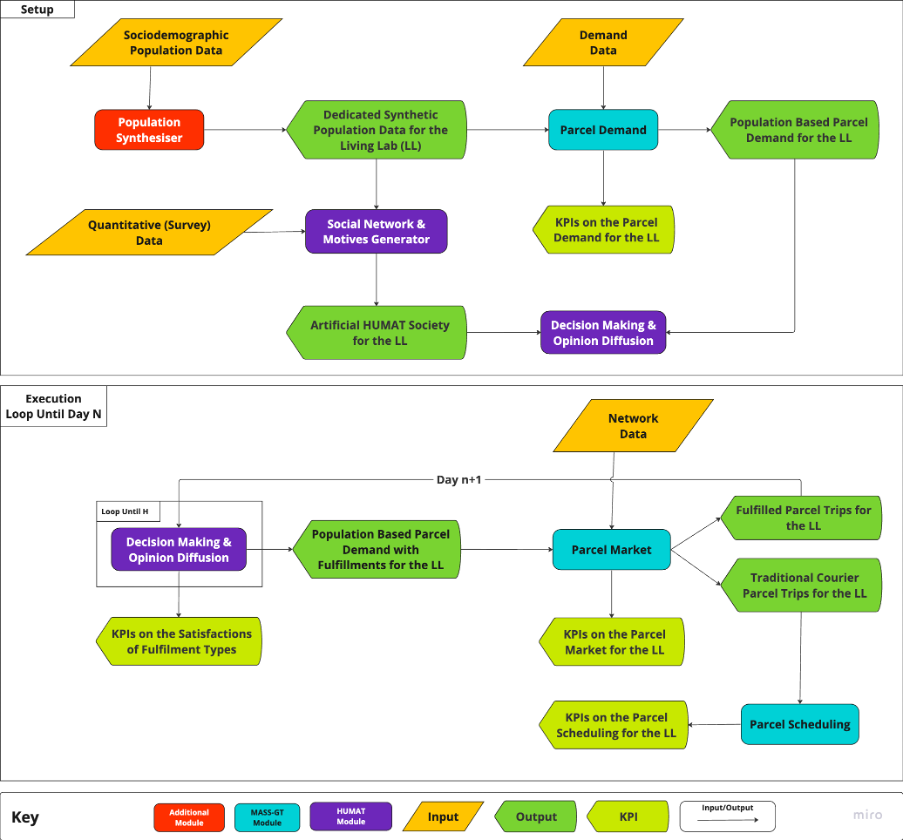}
\caption{Integrated generic modelling framework - all phases}
\label{fig:Modelling-Framework-Setup-Phase}
\end{figure}

Our integrated high-level modelling framework for LMD systems identifies the interconnections and complementarities between HUMAT and MASS-GT. It develops a logical structure between them that can ensure the transferability of the models for future digital social twinning applications. It is designed with modularity, standardized interfaces, and precise documentation, making it quickly be adopted for simulating the uptake of social innovation studies in various Wave 1 LLs in selected, relevant contexts (see Section 6). Due to the same properties, the adopted Wave 1 models are easily replicable for various Wave 2 LLs.      
The functional architecture consists of two main elements: (1) Setup, (2) Execution that are shown in Figure \ref{fig:Modelling-Framework-Setup-Phase}.

In the setup phase, the population synthesizer creates a synthetic population for the LLs using socio-demographic data like age, gender, and income, initially sourced from EUROSTAT and supplemented with LL-specific information. MASS-GT's Parcel Demand module then generates parcel demand and KPIs at the household level, with zonal-level demand linked to households based on network data.
HUMAT’s Social Network \& Motives Generator module creates an artificial society of diverse HUMAT agents, each based on sociodemographic data from the population synthesizer and needs and values from survey data in each LL. 
The Decision Making \& Opinion Diffusion module then uses these agents and parcel demand data to simulate social interactions and choice evaluations, calibrating behavioral preferences within the artificial population.

During the model execution phase, MASS-GT's Parcel Market module assigns parcel trips with their origins and destinations, generating KPIs for delivery services like crowdshipping and parcel locker (PL) facilities. Traditional courier parcels are then handled by the Parcel Scheduling module, where delivery tours are created.
HUMAT’s Decision Making \& Opinion Diffusion module determines which delivery services agents choose and facilitates communication among agents about these options. It also generates KPIs related to the satisfaction of individuals with the available delivery methods.   

%\subsection{Implementation of the Modelling Framework}
%\label{sec:Implementation-of-the-Modelling-Framework}

%HUMAT was initially created using NetLogo, while MASS-GT was implemented in Python. 
%HUMAT was later re-implemented in Python for improved integrity \cite{Gurcan2023}. 
%These components were deployed as Python modules to build the modelling framework and integrated into each calibrated use case (see Section \ref{sec:Case-Studies}).

\section{Case Study: Crowdshipping Service}
\label{sec:Crowdshipping-Service}

\begin{figure}[h!]
\centering
\includegraphics[width=0.82\textwidth]{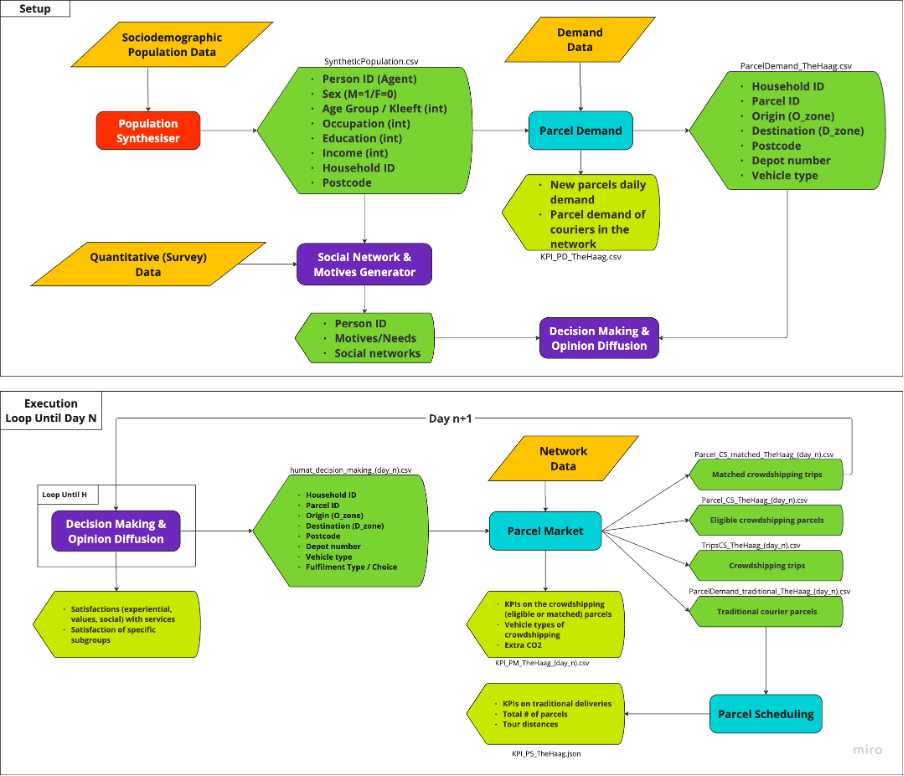}
\caption{Modelling framework calibrated for crowdshipping.}
\label{fig:Modelling-Framework-Crowdshipping}
\end{figure}

Crowdshipping, a novel LMD method, involves using everyday individuals as couriers to deliver packages from a central location to the final destination. This logistics model leverages local, non-professional couriers to complete the last leg of delivery, typically over short distances. It aims to increase efficiency, reduce costs, and speed up delivery by using existing travel routes of these crowdshippers.
In this use case, we calibrated the modelling framework given in Section \ref{sec:Modelling-Framework} to simulate consumer behavior in LMD crowdshipping services, using data collected from The Hague (see Figure \ref{fig:Modelling-Framework-Crowdshipping}).

As detailed in Section \ref{sec:Modelling-Framework},  the population synthesizer generates a synthetic population for both models. For The Hague, we use an existing synthetic population based on 4-digit postcodes (PC4). Open-source national statistics provide socio-demographic data to create a representative population for the Netherlands, including unique person and household IDs. Household data generates demand, while person data initializes HUMAT.  
Figure \ref{fig:Modelling-Framework-Crowdshipping} shows characteristics like sex, age group, education, and other socio-demographic factors.

Parcel demand is determined by analyzing household data and employment characteristics in different areas of the Netherlands\footref{ACM}\footnote{Centraal Bureau voor de Statistiek: Centraal bureau voor de statistiek. \url{https:
//www.cbs.nl/}, accessed on 31/08/2024.}. 
This data, along with courier market shares and depot locations, helps calculate demand for each courier, including pickup and delivery points. Consumer surveys on crowdshipping participation are used to allocate a portion of the demand. Finally, the parcel scheduling module organizes delivery tours for the parcels not handled by crowdshipping.
The population of consumers is generated in HUMAT using socio-demographic data from the population synthesizer, along with needs and values from consumer surveys. The agents then create their social network based on sociodemographic similarities.

\section{Case Study: Parcel Locker Service}
\label{sec:Parcel-Locker-Service}

\begin{figure}[h]
\centering
\includegraphics[width=0.82\textwidth]{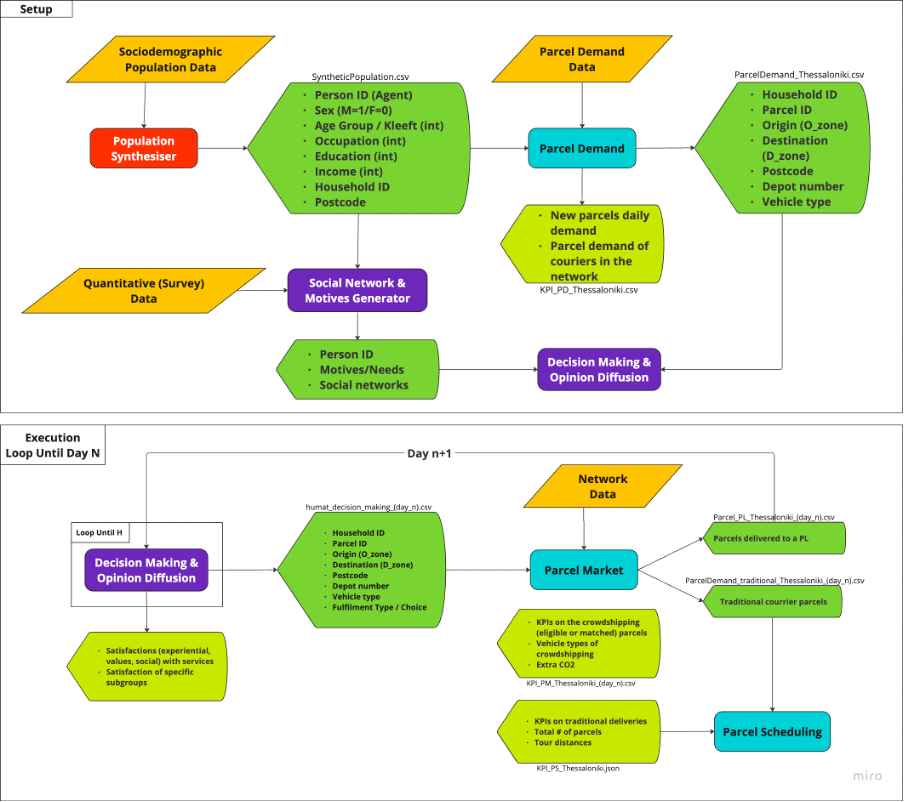}
\caption{Modelling framework calibrated for parcel locker service.}
\label{fig:Modelling-Framework-Parcel-Locker}
\end{figure}

Parcel locker services are an innovative component of last-mile delivery systems aimed at enhancing efficiency and convenience. These services involve the installation of secure, self-service lockers at various accessible locations, such as shopping centers, transportation hubs, residential areas, and workplaces. When a parcel is delivered, the consumer is notified and provided with a unique code to access the locker.
In this use case, we calibrated the modelling framework given in Section \ref{sec:Modelling-Framework} for modelling and simulating the consumer behavior in LMD parcel locker services  (see Figure \ref{fig:Modelling-Framework-Parcel-Locker}).

The population synthesizer generates a synthetic population using EU-SILC data \cite{Falck2024} for input into both models, reflecting the socio-demographic characteristics of the Thessaloniki LL (see Figure \ref{fig:Modelling-Framework-Parcel-Locker}). This includes person ID and zonal ID data.
To calibrate HUMAT, individual customer agents are created with socio-demographic traits from the synthesizer. Each agent is assigned needs based on a Thessaloniki LL survey, and a social network is formed based on socioeconomic similarities. Finally, agents choose their preferred delivery method, influencing the proportion of parcel locker deliveries.
For MASS-GT demand generation, household characteristics and income levels per zone generate demand, with HUMAT determining the share for parcel locker deliveries. Parcel distribution to lockers considers distance, facility capacity, and availability, calibrated for the LL in the parcel market module. Scheduling is then completed for both parcel locker and traditional courier deliveries.

%\subsection{Autonomous Delivery Robot Service}
%\label{sec:Autonomous-Delivery-Robot-Service}

%In this use case, we calibrated the modelling framework given in Section \ref{sec:Modelling-Framework} for modelling and simulating the consumer behavior in LMD autonomous delivery robot (ADV) services  (see Figure \ref{fig:Modelling-Framework-ADV}).

%\begin{figure}[h]
%\centering
%\includegraphics[width=0.95\textwidth]{figure/Calibrated_ADV_Model.png}
%\caption{Modelling framework calibrated for autonomous delivery robot service.}
%\label{fig:Modelling-Framework-ADV}
%\end{figure}

\section{Discussion and Conclusions}
\label{sec:Discussion-and-Conclusion}

This research presents a comprehensive framework that integrates agent-based modeling tools, specifically HUMAT and MASS-GT, to tackle the complexities of last-mile delivery (LMD) systems. 
The unified framework effectively simulates and evaluates urban freight dynamics across different cities, focusing on consumer behavior and delivery logistics, which are crucial for optimizing LMD strategies.
The case studies in Section \ref{sec:Crowdshipping-Service} and Section \ref{sec:Parcel-Locker-Service} demonstrate the framework's flexibility and adaptability to various urban settings and logistics scenarios, proving its value in planning and decision-making\footnote{The application and simulation results of URBANE project's case studies are reported in: \url{https://www.urbane-horizoneurope.eu/wp-content/uploads/2024/09/URBANE_D3.2_Modelling_Framework_and_Agent-Based_Models_v1.0-1.pdf}.}. 
These studies also reveal the significant impact of consumer behavior on LMD system efficiency, with the HUMAT model providing insights into how social networks and personal beliefs influence delivery choices.
Additionally, the research highlights the interdependence between freight transport agents and consumers, showing that consumer decisions can directly affect delivery service efficiency, congestion, and emissions. By incorporating these decision drivers, the framework enables more informed and sustainable transport strategies.

This generic modeling framework marks a significant advancement in urban logistics, offering a powerful tool for planners, policymakers, and logistics companies to improve delivery system efficiency and sustainability in urban areas. The success of the URBANE project, supported by detailed case studies, confirms the framework's practical applicability and potential for future customization to accommodate emerging technologies and challenges in urban logistics.
Future research should expand the range of delivery options and technologies within the framework, such as drones and AI-driven logistics, and explore the socio-economic impacts of different LMD strategies on diverse urban populations.
Overall, this research significantly contributes to computational social science by providing a tool that enhances the understanding of complex systems and supports the development of innovative, socially responsible solutions.

\section*{Acknowledgment}

The work reported here is part of the EU Horizon URBANE project with grant agreement No. 101069782.

\bibliographystyle{splncs03} 
\bibliography{references,bibliography}

\end{document}